# Controlling collective spontaneous emission with plasmonic waveguides

YING LI AND CHRISTOS ARGYROPOULOS*

*Department of Electrical and Computer Engineering, University of Nebraska-Lincoln, Lincoln, NE, 68588, USA*
*christos.argyropoulos@unl.edu*

**Abstract**: We demonstrate a plasmonic route to control the collective spontaneous emission of two-level quantum emitters. Superradiance and subradiance effects are observed over distances comparable to the operating wavelength inside plasmonic nanochannels. These plasmonic waveguides can provide an effective epsilon-near-zero operation in their cut-off frequency and Fabry-Pérot resonances at higher frequencies. The related plasmonic resonant modes are found to efficiently enhance the constructive (superradiance) or destructive (subradiance) interference between different quantum emitters located inside the waveguides. By increasing the number of emitters located in the elongated plasmonic channel, the superradiance effect is enhanced at the epsilon-near-zero operation, leading to a strong coherent increase in the collective spontaneous emission rate. In addition, the separation distance between neighboring emitters and their emission wavelengths can be changed to dynamically control the collective emission properties of the plasmonic system. It is envisioned that the dynamic modification between quantum superradiant and subradiant modes will find applications in quantum entanglement of qubits, low-threshold nanolasers and efficient sensors.

## 1. Introduction

Since the Purcell effect was proposed [1], the modification of the spontaneous emission rate emitted by quantum emitters matched to a resonant cavity has attracted much attention leading to several intriguing applications, such as efficient nanolaser sources [2], optical communication devices [3], DNA analysis [4], single-photon generation [5] and sensitive optical microscopy [6]. Purcell proved that the spontaneous emission decay rate is not an intrinsic property of the emitter but largely depends on the inhomogeneity of the environment. Therefore, resonant systems, such as nanocavities [7], photonic crystals [8], nanoshells [9], plasmonic waveguides [10] and nanoantennas [11], can be introduced to enhance this decay rate. However, one major limitation towards practical emission enhancement is that it is extremely difficult to boost the total emission of an ensemble of quantum emitters. In this scenario, each emitter usually needs to be accurately placed at a specific location in the resonating system, where large and homogeneous electric field distributions exist. This directly hinders the collective spontaneous emission rate enhancement, especially in the common scenario of a collection of quantum emitters arbitrarily located inside a resonating system.

This collective spontaneous emission response, also known as superradiance, can improve the directivity and coherence of the total emitted radiation by an ensemble of quantum emitters. Considerable research efforts have been devoted to the study of superradiance because of its interesting potential applications in quantum communications [12,13], narrow linewidth lasers [14], atom lasers [15] and thermal emitters [16]. The phenomenon of superradiance was originally proposed by Dicke [17], who demonstrated that the radiation intensity emitted by $N$ atoms placed in subwavelength distances was proportional to $N^2$ instead of the usual $N$. This phenomenon is based on the constructive interference between emitted waves and has been investigated by using quantum emitters coupled to a variety of photonic environments including microcavities [18,19], metal interfaces [20], plasmonic waveguides [21,22] and left-handed media [23]. However, this effect can only occur when the neighboring quantum emitters are separated by a small fraction of the emitted radiation wavelength [12], which limits its practical applications. In addition, the counterpart mechanism of superradiance is subradiance. It is a destructive interference effect leading to suppressed emission from a collection of active particles, such as atoms [24–27], ions [28] and quantum dots (QDs) [21,29]. Subradiant states are difficult to be created, similar to superradiant states, and, moreover, are not easy to be observed [27]. In this case, the collective emission rate is inhibited and equal to zero because the quantum emitters experience destructive interference in their collective state. Through coherent manipulations of collective atomic states, superradiant modes can be transformed into subradiant modes and vice versa [30,31], with possible applications on optical quantum memories [32], quantum computers [33,34] and nanolasers [35].

In recent years, realistic metamaterials with effective epsilon-near-zero (ENZ) permittivity response have generated increased interest, especially due to their peculiar transmission properties that provide, in principle, infinite phase velocity combined with anomalous impedance-matching. The ENZ response has been theoretically predicted [36–39] and experimentally verified [40] using narrow plasmonic waveguides operating at their cut-off wavelength. Uniform phase distribution and large field enhancement is obtained inside the channels of these narrow waveguides. This anomalous quasi-static response is independent of the channel's length and shape and has been used to squeeze and tunnel light [39], enhance fluorescence [41,42], boost optical bistability [43], excite temporal solitons [44] and obtain giant second harmonic generation [45]. Large and uniform local density of states (LDOS) can be achieved inside the plasmonic waveguides at the ENZ cut-off wavelength and these are ideal conditions to increase the collective spontaneous emission rate of several emitters. Additionally, Fabry-Pérot (FP) resonances are dominant in larger frequencies above the ENZ cut-off frequency [43]. These higher-order resonances can also enhance LDOS of emitters placed in particular locations of the plasmonic channel.

In this work, we demonstrate a way to obtain different collective spontaneous emission effects, such as superradiance and subradiance, excited by a collection of quantum emitters placed inside plasmonic channels. ENZ operation is obtained at the cut-off wavelength of these plasmonic waveguides and FP resonances are found at lower wavelengths. Classical electromagnetic calculations are used to compute the Purcell enhancement and radiative efficiency of a single emitter and a pair of two-level quantum emitters embedded inside the plasmonic waveguide at ENZ and FP resonances. The utilized two-level emitters are characterized by the ground state and the excited state. They modeled using the point-dipole approximation, assuming weak excitation (no saturation) and operation in the weak coupling quantum regime [46]. Strong superradiance is obtained at the ENZ wavelength that is independent of the emitters' distance. It can be achieved without the usual constraint of subwavelength distance between emitters and can be obtained even at emitter distances on the order of wavelength. In addition, both superradiant and subradiant modes exist at higher-order FP resonances found in lower wavelengths compared to ENZ. It is demonstrated that these collective responses can be dynamically controlled by changing the emitters' separation distance, location, and frequency of operation. We also consider the collective emission of $N$ emitters uniformly located inside the plasmonic channel at the ENZ wavelength. In this case, the collective decay factor is amplified by almost $N$ times compared to a single emitter's decay rate. This is typical response of systems exhibiting superradiance [29,47]. We also theoretically compute the power time-dependent decay curves for different number of emitters. These results directly correspond to power lifetime measurements that can be obtained in an experimental verification of the proposed plasmonic superradiance effect. Finally, we provide insights on the time-dynamics of the proposed collective emission effects. The proposed quantum plasmonic system is envisioned to have several applications in quantum communication and computing systems on a chip, such as low-threshold nanolasers [35], quantum memories [32], and ultrasensitive optical sensors [48,49].

## 2. Optical response of plasmonic waveguides

The geometry of the proposed plasmonic grating unit cell is shown in Fig. 1. A narrow rectangular slit is carved in a silver (Ag) screen, whose permittivity dispersion follows previously derived experimental data [50]. The slit has width $w$ = 200 nm, height $t$ = 40 nm and length $l$ = 500 nm. It is loaded with silica with relative permittivity of $\varepsilon$ = 2.2. This structure was originally introduced before [43] for nonlinear applications and can sustain ENZ and FP resonances. The width $w$ is designed to tailor the cut-off wavelength of the dominant quasi-$TE_{10}$ mode propagating inside each slit. At this frequency point, the plasmonic waveguide behaves as an effective ENZ medium and this anomalous impedance-matching phenomenon leads to total transmission combined with large field enhancement and uniform phase inside each slit [39]. This effect is independent of the grating's periodicity or channels' length $l$ [43]. In this work, the grating period was chosen to be $a$ = 400 nm and $b$ = 400 nm but similar effects are expected from an isolated plasmonic slit.

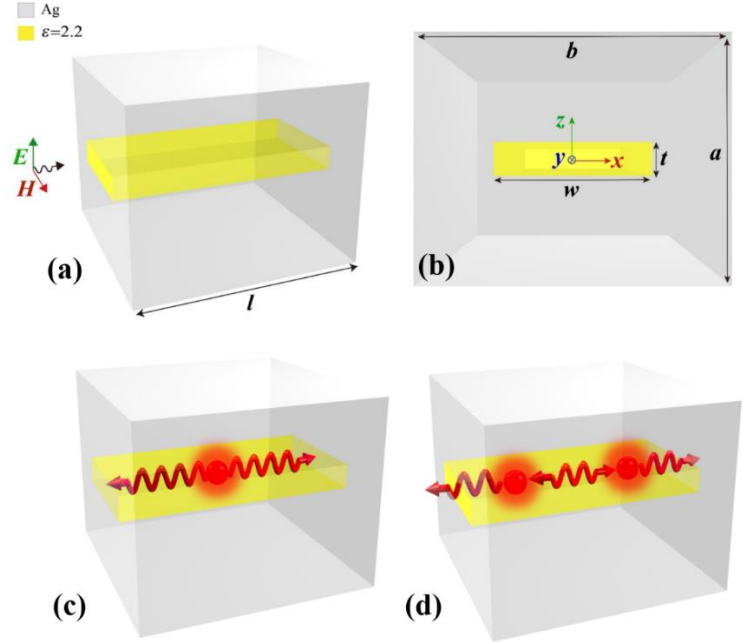

Fig. 1. (Color online) Geometry of the silver plasmonic waveguide loaded with glass. A rectangular slit is carved in a silver screen. (a) The device is excited by a plane wave impinging at normal incidence and the transmittance is calculated. (b) Cross-sectional view of the unit cell geometry. (c) A single emitter is placed inside the plasmonic channel to calculate the spontaneous emission rate. (d) A pair of two-level quantum emitters is placed inside the plasmonic channel to investigate the effect of superradiance and subradiance.

Since the slits occupy a very small area on the grating surface, the incident waves are mostly reflected at the first interface. However, optical transmission can occur around the cut-off frequency and at higher FP frequencies leading to minimum reflection and maximum transmission. The plasmonic grating is illuminated by a normal incident $z$-polarized plane wave shown in Fig. 1(a). The computed transmittance is shown in Fig. 2(a) as a function of the incident radiation wavelength. The transmission peak at $\lambda$ = 1012 nm corresponds to the cut-off frequency of the dominant quasi-$TE_{10}$ mode. The field distribution normalized to the incident wave is homogeneous and enhanced along the channel at this wavelength [Fig. 2(b)]. As it was expected, the plasmonic waveguide effectively behaves as ENZ material at the cut-off wavelength. For longer wavelengths above the cut-off value, the incident wave is totally reflected by the structure and the transmission is zero. Below the cutoff wavelength, an additional transmission peak appears at $\lambda = 922$ nm corresponding to the first-order FP resonance. The field enhancement distribution at the FP resonance is shown in Fig. 2(b) where a typical standing wave pattern is obtained. The grating is excited by a plane wave impinging at normal incidence but the ENZ operation will be unaffected even at oblique incidence. The ENZ transmittance is 50% due to increased losses at near-infrared (near-IR) coming from the silver waveguide walls. The ENZ peak is slightly lower compared to the FP resonance peak due to the uniform field distribution at the ENZ wavelength, making the optical absorption from the silver walls more effective. The ENZ transmission can be further increased, in case we reduce the waveguide's length, without affecting the ENZ performance.

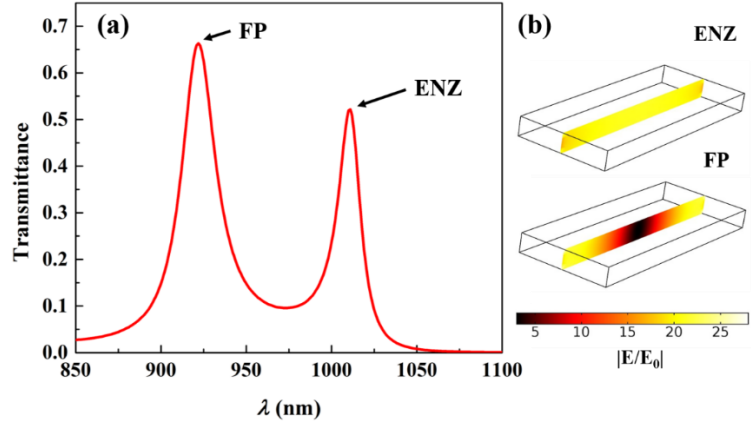

Fig. 2. (Color online) (a) Transmittance of the plasmonic channel as a function of the incident wavelength. (b) The total electric field enhancement distribution in the channel's *yz*-plane operating at the ENZ and FP resonant wavelengths.

## 3. Spontaneous emission of single emitter in plasmonic waveguides

We compute the spontaneous emission rate excited by a single two-level quantum emitter embedded in the narrow plasmonic waveguide shown in Fig. 1(c). Assuming weak excitation (no saturation) and operation in the weak coupling regime, the emitter can be modeled using the point-dipole approximation [46]. The spontaneous emission decay rate at position $\mathbf{r} = \mathbf{r}_0$ can be expressed as [51,52]:

$$\gamma_{sp} = \frac{\pi \omega_0}{3\hbar \varepsilon_0} |\mathbf{\mu}|^2 \rho(\mathbf{r}_0, \omega_0), \qquad (1)$$

where $\omega_0$ is the angular emission frequency, $\mathbf{\mu}$ is the emitter's transition dipole moment, and $\rho(\mathbf{r}_0, \omega_0)$ is the LDOS corresponding to the excited electromagnetic mode at a given frequency and location $\mathbf{r}_0$. The LDOS can be calculated by using the system's dyadic Green's function $\mathbf{G}$:

$$\rho(\mathbf{r}_0, \omega_0) = \frac{6\omega_0}{\pi c^2} \left[ \hat{\mathbf{n}} \cdot \text{Im}\{\mathbf{G}(\mathbf{r}, \mathbf{r}_0)\} \cdot \hat{\mathbf{n}} \right], \qquad (2)$$

where $\hat{\mathbf{n}}$ is the unit vector in the direction of the dipole moment ( $\mathbf{\mu} = \mu \hat{\mathbf{n}}$ ). $\mathbf{G}(\mathbf{r}, \mathbf{r}_0)$ is defined by the electric field $\mathbf{E}$ at the point $\mathbf{r}$ induced by an electric dipole at the source point $\mathbf{r}' = \mathbf{r}_0$:

$$\mathbf{E}(\mathbf{r}) = \omega_0^2 \mu_0 \mu \mathbf{G}(\mathbf{r},\mathbf{r}')\boldsymbol{\mu}, \tag{3}$$

with $\mu_0$ the permeability of free space and $\mu$ the relative permeability of the surrounding space.

In plasmonic systems, the total spontaneous emission is decomposed into its radiative (energy transferred into the environment) and non-radiative (associated with system losses) contributions [52]: $\gamma_{sp} = \gamma_r + \gamma_{nr}$. Using full-wave three-dimensional simulations based on the finite element method (COMSOL Multiphysics), we calculate the LDOS excited by a single emitter at the *yz*-plane of the plasmonic waveguide channel. Next, the total spontaneous emission rate $\gamma_{sp}$ is derived using Eq. (1) assuming an emitter with a transition dipole moment $\boldsymbol{\mu}$ equal to 1 [C·m]. The non-radiative rate $\gamma_{nr}$ can also be computed with COMSOL [52]. The computed total spontaneous emission and non-radiative rates lead to the calculation of a very important quantity, named quantum yield *QY*, which is defined as the ratio $QY = \gamma_r / \gamma_{sp}$ and reflects the radiative emission efficiency [52]. It quantitatively describes how much radiation can escape from the plasmonic system to the surrounding space and is nearly independent of emitter's intrinsic quantum yield [53]. The emitter is assumed to be vertically oriented (*z*-axis) that guarantees a maximum coupling with both ENZ and FP modes of the plasmonic structure. In free space, a point dipole emitter has LDOS equal to $\rho = \omega^2/(\pi^2 c^3)$ and corresponding radiative rate $\gamma_r^0 = \omega^3 |\boldsymbol{\mu}|^2 / (3\hbar\pi\varepsilon_0 c^3)$. The total spontaneous emission rate at free space is given by $\gamma_{sp}^0 = \gamma_r^0 / QY_0$, where $QY_0$ is the free-space quantum yield of the used emitters. Here, this property is chosen to be $QY_0 = 0.2$, typical value of fluorescence emitters [54].

The normalized to free space spontaneous decay rate $\gamma_{sp}/\gamma_{sp}^0$ and quantum yield *QY* distributions are plotted in Fig. 3 at the ENZ and FP resonances. These maps were calculated by varying the emitter position on a discrete 4x50 grid placed at the channel's *yz*-plane. The spontaneous emission rate is boosted up to 200 times and is spatially uniform along the channel's length at the ENZ wavelength [Fig. 3(a)]. The radiative emission efficiency, quantified by the *QY*, reaches high values of 0.7 and its distribution is also uniform [Fig. 3(b)]. On the contrary, at the FP resonance, both emission enhancement and *QY* are largely dependent on the location of the dipole emitter along the plasmonic channel, as it is shown in Figs. 3(c) and (d), respectively. This is consistent with the electric field standing wave distribution shown before in Fig. 2(b). Additionally, the maximum spontaneous emission rate enhancement is lower at the FP wavelength compared to the ENZ response. Furthermore, the plasmonic waveguide demonstrates a highly directional radiation pattern at the ENZ resonance, which is computed and shown in Fig. 4, assuming that the dipole emitter is vertically oriented inside the nanochannel. The directional radiation pattern is independent of the emitter's position inside the waveguide at the ENZ wavelength. Directional radiation is very important for designing the future integrated optical communication nanodevices.

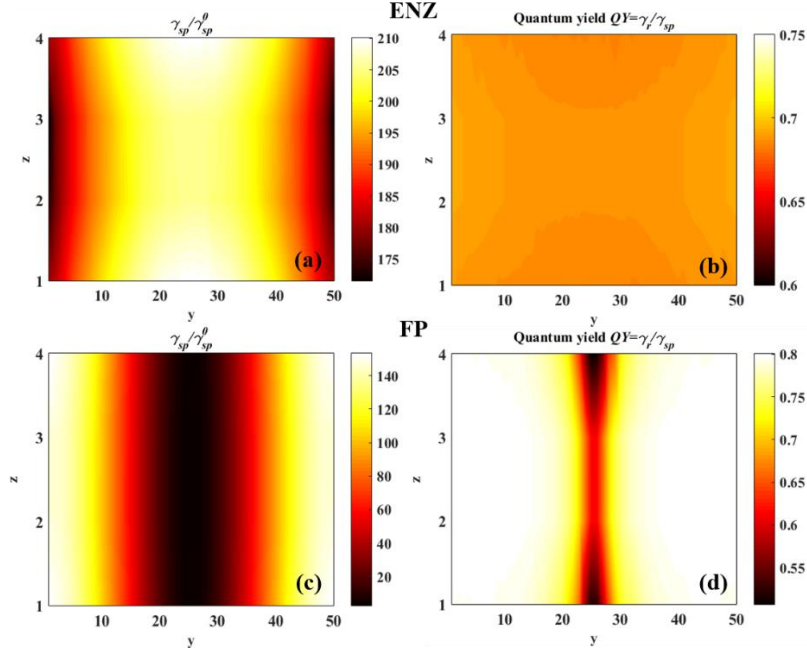

Fig. 3. (Color online) (a, c) Normalized spontaneous decay rate $\gamma_{sp}/\gamma_{sp}^0$ and (b, d) radiative quantum yield $QY = \gamma_r/\gamma_{sp}$ distributions excited by one emitter located at the channel's $yz$-plane and emitting at the (a, b) ENZ and (c, d) FP resonant wavelengths. The $y$ and $z$ axes in these figures scale with the simulated 4×50 discrete grid used to place the emitter at the channel's $yz$-plane.

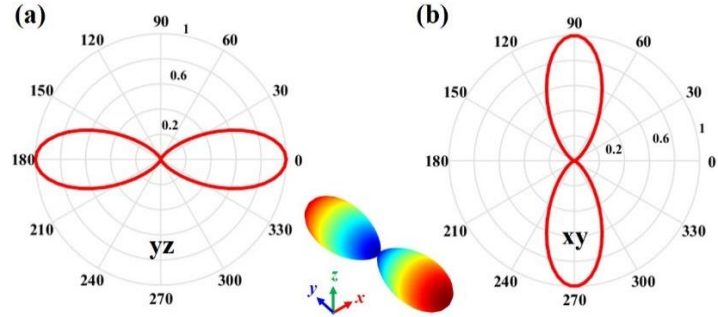

Fig. 4. (Color online) (a) Normalized radiation pattern in the $yz$-plane and (b) in the $xy$-plane for a dipole emitter placed inside the plasmonic waveguide channel at the ENZ resonance. Inset: Three-dimensional radiation pattern.

## 4. Collective spontaneous emission of multiple emitters in plasmonic waveguides

Next, we consider the collective spontaneous emission properties and the coherent interactions from a pair of two-level quantum emitters to investigate the effect of superradiance and subradiance. They are placed at the center of the plasmonic waveguide channel along the $y$-axis and their distance can change [see Fig. 1(d)]. Again the point-dipole approximation is used in these calculations, which is valid for small emitters operating at the
8

weak coupling regime. In addition, the emitters are oscillating in-phase and are made of the same material. This is a typical behavior of molecules inside active bulky media, such as fluorescence materials. We assume that the location of the first emitter *i* with dipole moment $\boldsymbol{\mu}_i$ is fixed at position $\mathbf{r}_i$ (near the channel's edge). The second emitter *j* with identical dipole moment $\boldsymbol{\mu}_j$ is placed at position $\mathbf{r}_j$, which varies along the *y*-axis. Both emitters oscillate in phase and are separated by distance *d*. In this case, the non-local density of states (NLDOS) is computed in order to define the resultant density of states due to interference caused by the second emitter to the first one and inverse. The total decay rate based on the NLDOS is [21]:

$$\gamma_{ij} = \frac{2\omega_0^2}{\varepsilon_0 \hbar c^2} \text{Im}[\boldsymbol{\mu}_i^* \cdot \mathbf{G}(\mathbf{r}_i, \mathbf{r}_j) \cdot \boldsymbol{\mu}_j], \qquad (4)$$

where $\boldsymbol{\mu}_i^*$ is the complex conjugate of the transition dipole moment of emitter *i*. Therefore, $\gamma_{12}$ represents the contribution to the decay rate of emitter 1 at position $\mathbf{r}_1$ due to interference caused by emitter 2 located at position $\mathbf{r}_2$. Hence, a normalized decay factor $\gamma$ is defined to quantify the modification of the collective decay rate due to interference from another emitter:

$$\gamma = \frac{\gamma_{11} + \gamma_{12} + \gamma_{22} + \gamma_{21}}{\gamma_{11} + \gamma_{22}}. \qquad (5)$$

We numerically calculate the normalized total decay factor $\gamma$. The result is plotted in Fig. 5(a) versus the two emitters' separation distance *d* at the ENZ ($\lambda$ = 1012 nm/ black curve) and FP ($\lambda$ = 922 nm/red curve) resonances, respectively. The green dashed line refers to the end of the plasmonic channel. Outside of the slit is assumed to be free space and $\gamma$ naturally converges to one after a small distance from the waveguide's edge. In this case, the second emitter is located outside of the waveguide and the coupling between the emitters due to the plasmonic waveguide modes ceases to exist. Superradiant (subradiant) modes are excited when the normalized decay factor $\gamma$ is larger (smaller) than one due to constructive (destructive) interference between the emitters operating at ENZ and FP resonances. The decay factor $\gamma$ is also calculated when the two emitters are uncoupled and placed in free space [blue curve in Fig. 5(a)]. Note that the normalized total decay factor curves tend to the same limit: $\gamma = 2$ (Dicke's perfect superradiance) or $\gamma = 1$ (free space uncoupled emission) at $d \to 0$ or $d \to \infty$, respectively, for different operation frequencies and with and without the plasmonic waveguide.

Interestingly, perfect superradiance ($\gamma = 2$) is achieved along the entire waveguide's length only at the ENZ operation. The length of the waveguide is comparable to the effective wavelength $\left(\lambda_{\text{eff}} = \lambda/\sqrt{\varepsilon} \approx 670 nm\right)$ of the emitted radiation. Note that in free space [blue curve in Fig. 5(a)], superradiance can normally be achieved only if the emitters are very closely packed to each other, confined in highly subwavelength regions $\left(d \approx \lambda_{\text{eff}}/10\right)$ [12]. This detrimental property severely limits its practical applications. However, subwavelength distance is not needed in order to obtain superradiance when emitters are placed inside plasmonic channels operating at the cut-off (ENZ) wavelength. In addition, the ENZ superradiance emission is directional, as it was shown in Fig. 4. Strong superradiance can also be obtained when the channel's length is increased, which is not going to affect the ENZ response. Therefore, ENZ can extend this interesting effect to regions comparable and even larger to the wavelength. In this case, more emitters can be incorporated inside the nanochannel, leading to much stronger superradiant emission that is strongly directional in space and time [12]. The strong and uniform field enhancement at ENZ resonance [Fig. 2(b)] is responsible for this effect. Hence, the ENZ mode leads to perfectly coherent interactions between separate emitters.

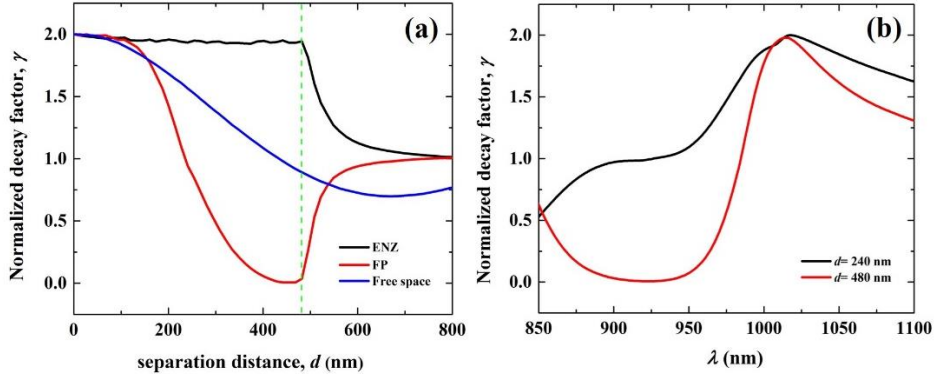

Fig. 5. (Color online) (a) Normalized collective decay factor $\gamma$ versus the separation distance $d$ at the ENZ ($\lambda = 1012$nm/black curve) and FP ($\lambda = 922$nm/red curve) resonances. The blue curve shows the variation of $\gamma$ when the two emitters are placed in free space. The green dashed line depicts the end of the plasmonic channel. (b) Normalized decay factor $\gamma$ versus the excited wavelength $\lambda$ for two different emitter separation distances: $d = 240$ nm (black curve) and $d = 480$ nm (red curve).

On the contrary, at the FP resonance, the perfect superradiant mode ($\gamma = 2$), when the emitters are located very close, gradually changes to the perfect subradiant mode ($\gamma = 0$), when the distance between the emitters is increased and becomes equal to the channel's length. Different positions of the second emitter provide different phase shifts between the two coupled emitters, which originally oscillate in phase. In principle, subradiant systems can store photons and the information encrypted onto them for a long time leading to the design of new types of quantum memories [33,34]. Figure 5(b) demonstrates the relationships between the normalized collective decay factor $\gamma$ versus the emitted wavelength with fixed separation distances between the emitters: $d = 240$ nm (black curve) and $d = 480$ nm (red curve). Superradiance ($\gamma = 2$) is obtained at the ENZ wavelength ($\lambda = 1012$ nm), independent of the emitter's separation distance, consistent with the results in Fig. 5(a). On the contrary, at the FP resonance ($\lambda = 922$ nm), the collective emission properties of the plasmonic system can be dynamically controlled and changed from subradiant state ($\gamma = 0$) to free-space state ($\gamma = 1$) depending on the emitter's distance. Note that subradiance can only be observed at fixed emitter's distance operating at the FP resonance mode, consistent with the standing wave distribution shown in Fig. 2(b).

We also compute the collective emission enhancement from an arbitrary large number of quantum emitters $N$ uniformly distributed in the narrow waveguide. In the case of two quantum emitters placed in the channel operating at the ENZ resonance, the normalized decay factor given by Eq. (5) has been calculated to be $\gamma = 2$ (Dicke's perfect superradiance). Similarly, when $N$ quantum emitters are placed in the plasmonic ENZ system, they will exhibit perfect superradiance and their normalized collective decay factor can reach the maximum value $\gamma = N$, i.e., it will be equal to the total number of emitters. This can be computed by the generalized version of Eq. (5) for $N$ coherently interacting emitters. Note that $\gamma$ is analogous to the total electric field $\gamma \propto E$. The emitted radiation intensity is proportional to the square of the electric field and, as a result, it will be analogous to $N^2$, as it was predicted by Dicke [17]. We numerically calculate this normalized total decay factor $\gamma$ excited by $N = 100$ emitters at the channel's *yz*-plane at the ENZ resonance (see Fig. 6). Uniform and strong enhancement of the total collective emission is obtained ($\gamma$ is very close to *N* inside the nanochannels), which means that all emitters in the nanochannel are involved in the collaborative superradiance effect independent of their positions. Hence, we can pack as many emitters as possible inside the waveguides and we can obtain a giant increase in the total spontaneous emission rate at the ENZ resonance of this plasmonic configuration. The total superradiance of the plasmonic system is only limited by the number of emitters (active molecules) placed inside the nanochannel. Note that the superradiant state excited by $N = 100$ emitters also has high directionality in the far-field (same radiation pattern as Fig. 4) in the plasmonic ENZ system, which is a basic feature of the coherent superradiance effect [55].

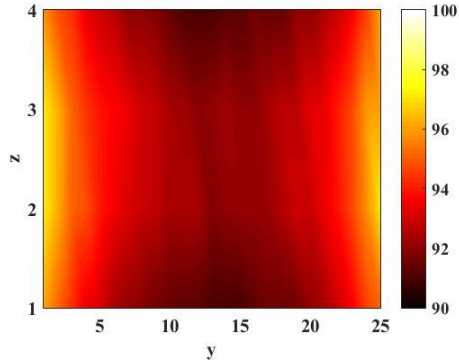

Fig. 6. (Color online) Distribution of normalized collective decay factor $\gamma$ excited by $N = 100$ emitters at the ENZ resonant wavelength. The emitters are uniformly located at the channel's *yz*-plane. The *y* and *z* axes scale with the simulated 4×25 discrete grid used to place $N = 100$ emitters along the channel's *yz*-plane. Each position on the discrete grid corresponds to one emitter.

## 5. Time-dependent lifetime decays and time-dynamic response

Finally, we calculate the time-dependent lifetime decay curves when one, two and $N = 100$ emitters are placed inside the plasmonic waveguide nanochannel discussed before. The emitted power from an emitter at point **r** is proportional to the excitation field, the radiative rate and the total spontaneous emission. It is given from this formula [52,54]:

$$\mathbf{W}_{\hat{\mathbf{n}}}(\mathbf{r},t) \propto \left|\mathbf{E}_{ex}(\mathbf{r})\cdot\hat{\mathbf{n}}\right|^2 \gamma_r(\mathbf{r}) \exp\left[-\gamma_{sp}(\mathbf{r},\hat{\mathbf{n}})t\right], \quad (6)$$

where $\mathbf{E}_{ex}(\mathbf{r})$ is the field distribution at the excitation frequency. In all our simulations, the same excitation field distribution is assumed, no matter how many emitters are placed inside the channel. In addition, the orientation of the dipole moment for each emitter is assumed to be uniform. Only emitters oriented along the *z*-axis significantly contribute to the spontaneous emission, which can be written as [52]:

$$\gamma_{sp}(\mathbf{r},\hat{\mathbf{n}}) = \gamma_{sp}(\mathbf{r},\hat{\mathbf{z}})\cos^2\theta, \quad (7)$$

where $\theta$ is the angle between the emitter orientation $\hat{\mathbf{n}}$ and the *z*-axis. Therefore, the average emitted power over all directions can be expressed by [52]:

$$\bar{\mathbf{W}}(\mathbf{r},t) = \frac{1}{4\pi} \int_0^{2\pi} \int_0^{\pi} \mathbf{W}_{\hat{\mathbf{n}}}(\mathbf{r},t) \sin\theta d\theta d\phi$$
$$\propto \frac{1}{2} |\mathbf{E}_{ex}(\mathbf{r})|^2 \gamma_r(\mathbf{r}) \int_0^{\pi} \cos^2\theta \exp\left[-\gamma_{sp}(\mathbf{r},\hat{\mathbf{z}})\cos^2\theta t\right] \sin\theta d\theta. \tag{8}$$

In our calculations, we assumed the intrinsic lifetime of the emitter to be equal to $1/\gamma_{sp}^0 = 600$ ns, consistent with the experimental parameters obtained for Ru dyes [54]. By summing up the contribution of each emitter at the ENZ wavelength, the normalized emission curves are plotted as a function of time in Fig. 7 when the plasmonic system is excited by one emitter (red curve), two emitters (blue curve), and $N = 100$ emitters (black curve) placed inside the waveguide nanochannel. The strong enhancement in the total spontaneous emission rate due to superradiance can be directly seen in these time resolved emission curves. The emission is drastically decreased in time with an increase in the emitters' number. Interestingly, the substantial faster lifetime decay rates seen in Fig. 7 are similar to experimental superradiance results obtained from an ensemble of QDs [29]. This ultrafast response is ideal condition to create ultrafast coherent light sources for new optical communication networks [54].

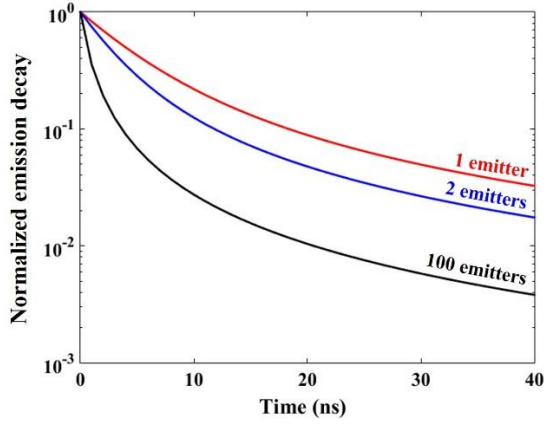

Fig. 7. Normalized time-dependent emission decay curves for the plasmonic waveguide excited by one emitter (red curve), two emitters (blue curve) and 100 emitters (black curve).

We would also like to note that ENZ plasmonic waveguides are characterized by slow-group velocities [44]. This effect may have an impact on the time dynamics of superradiance and is further analyzed. Three timescales are involved in the observation of collective behavior in spontaneous emission [55]: the spontaneous decay time of a single emitter $T_1 = 1/\gamma_{sp}$, the transit time of a photon travelling through the plasmonic waveguide $T_t = l/v_g$, and the characteristic decay time of collective processes (superradiance) given by [56] $T_{SR} \approx 2\varepsilon\hbar v_g / (n|\mathbf{\mu}|^2 \omega_0 l)$, where $v_g$ is the group velocity of a photon inside the waveguide, $n = N/(Al)$ is the concentration of emitters per unit volume and $\mathbf{\mu}$ is the electric dipole moment. In our work, we assumed that $N = 100$ emitters are embedded in the plasmonic waveguide channel with length $l$ and slit area $A$ ($l \gg \sqrt{A}$). In this plasmonic waveguide configuration, $v_g$ can be smaller than the velocity of light in free-space ($c$) at the ENZ resonance frequency. It was shown before that slow group velocity values of $v_g = c/26$ can be achieved close to the ENZ wavelength [44] for a plasmonic waveguide with similar dimensions. The proposed system is plasmonic, i.e. dominated by strong optical losses, and as a result the obtained slow group velocity is larger compared to other slow or stopped light photonic systems based on dielectric configurations, such as photonic crystals [57].

When the nanochannel is loaded with emitters with long intrinsic lifetimes $1/\gamma_{sp}^0 = 600$ ns (Ru dyes [54]), we can compute the relevant timescales at the ENZ resonance: $T_1 \approx 3\,\text{ns}$, $T_t \approx 43\,\text{fs}$ and $T_{SR} \approx 33.2\,\text{ps}$, which satisfy the condition $T_t \ll T_{SR} \ll T_1$. This scenario is ideal for observing pure superradiant emission with directional peak intensities several orders of magnitude higher than the intensity of one-emitter spontaneous emission [55]. However, if we choose to load the nanochannel with emitters characterized by short spontaneous lifetimes $1/\gamma_{sp}^0 = 3$ ns and larger electric dipole moments (DCM dyes [58] or QDs), the new time scales become $T_1 \approx 15\,\text{ps}$ and $T_{SR} \approx 166\,\text{fs}$ at the ENZ wavelength, leading to the modified time-dynamic condition $T_t \approx T_{SR} \ll T_1$. When this condition is satisfied, some of the collective radiated energy re-enters the plasmonic waveguide and pure superradiance is not possible to occur. In this case, the collective radiation emission takes the form of trains of pulses with decreased peak value, a phenomenon also known as oscillatory superradiance [55]. Hence, slow emitters need to be used in the potential experimental verification of the proposed plasmonic system in order to observe a pure superradiant response.

## 6. Conclusions

We propose an efficient way to control the collective spontaneous emission of photons from emitters. We numerically compute superradiant and subradiant modes excited by collections of quantum emitters placed inside plasmonic waveguides. The waveguide channels have an ENZ resonant response at their cut-off wavelength, where uniform phase distribution and large field enhancement is present. The spontaneous emission and quantum efficiency excited by a single emitter in the waveguide nanochannel is accurately computed. The collective emission properties of a pair of emitters is also analyzed. It is found that a directional superradiant state exists at the ENZ wavelength that is independent of the emitters' distance. Hence, superradiance is not limited to subwavelength distances between emitters and can be extended to distances comparable to the wavelength with this configuration. In addition, strong subradiance was observed at the FP resonance of the plasmonic waveguide, which inhibits the collective spontaneous emission properties again in distances comparable to the radiation's wavelength. By increasing the number of emitters located in the elongated plasmonic waveguide, the superradiance is further enhanced at the ENZ leading to an ultrafast response in the total emission rate. Note that all our simulations are conducted in the weak coupling regime and the superradiance and subradiance come from the waveguide modes and not from strong coupling effects. Moreover, dynamic tunability between superradiant and subradiant modes can be achieved, when the emitters operate at the different ENZ and FP resonant wavelengths. Finally, we would like to stress that enhanced superradiance at ENZ was also demonstrated in [22] but for an alternative waveguide structure and using a different theoretical approach. In this work, we present both superradiance and subradiance with a different plasmonic grating standing in free-space, making the concept much broader and more attractive for a variety of optical applications. Moreover, the presented theoretical analysis is extended to the calculation of the far-field distribution, the time-dependent lifetime decay and, even more importantly, the time-dynamic response. The current results are expected to pave the way to the experimental realization of this interesting quantum plasmonic concept. In addition, superradiant states can be converted into subradiant states and the inverse with the proposed quantum plasmonic system just by changing the emitters operating frequency. Such an ability to control and enhance or inhibit the total spontaneous emission rate can have fundamental implications in quantum communication and computing systems [32–34], low-threshold nanolasers [35], ultrasensitive optical sensors [48,49] and new solar cell designs. Our findings can also be applied to other quantum processes, such as the efficient generation and control of long range quantum entanglement between qubits [59,60].

**Funding**

This work was partially supported by the Office of Research and Economic Development at University of Nebraska-Lincoln and the National Science Foundation (NSF) through the Nebraska Materials Research Science and Engineering Center (MRSEC) (grant No. DMR-1420645).